\documentclass[conference]{IEEEtran}
\IEEEoverridecommandlockouts
\usepackage{cite}
\usepackage{amsmath,amssymb,amsfonts}
\usepackage{algorithmic}
\usepackage{graphicx}
\usepackage{textcomp}
\usepackage{xcolor}
\usepackage{tipa}
\usepackage{url}
\def\BibTeX{{\rm B\kern-.05em{\sc i\kern-.025em b}\kern-.08em
    T\kern-.1667em\lower.7ex\hbox{E}\kern-.125emX}}
\begin{document}

\title{Why Do You Say It Like That?\\ A Phoneme-Level Framework for Explainable Speech Deepfake Detection}

\title{Why Do You Say It Like That?\\ A Phoneme-level Framework for Explainable Speech Deepfake Detection\\

\thanks{This work was supported by the COMPROMIS project (ANR22-PECY-0011) funded by a French government grant managed by the Agence Nationale de la Recherche under the France 2030 program.}}


\author{
\IEEEauthorblockN{
Anna Taylor$^{1}$,
Michele Panariello$^{1}$,
Massimiliano Todisco$^{1}$,
Chiara Galdi$^{1}$,
Nicholas Evans$^{1}$,
Driss Matrouf$^{2}$
}
\IEEEauthorblockA{
$^{1}$EURECOM, Sophia Antipolis, France \\
$^{2}$Laboratoire Informatique d’Avignon, Avignon Universit\'e, France \\
\texttt{$^{1}${firstname.lastname}@eurecom.fr} \\
\texttt{$^{2}$driss.matrouf@univ-avignon.fr}
}
}
\maketitle

\begin{abstract}
As the accuracy of speech deepfake detection improves with the use of self-supervised representations such as wav2vec 2.0 and HuBERT, understanding why the speech is classified as bona fide or deepfake remains an open challenge. In pursuit of more trustworthy and interpretable AI, we introduce a phoneme-level analysis framework that connects model predictions to phonetic units. Our post-hoc explainability method is generally applicable to a variety of speech deepfake detection systems based on convolutional neural networks since it leverages Gradient-weighted Class Activation Mapping in conjunction with speech recognition to generate saliency maps aligned with phonemes and pauses. This pipeline reveals statistically significant attack- and speaker-dependent phonetic cues associated with spoofed speech in terms that humans can understand. Experiments using ASVspoof~5 show comparable detection performance to similar architectures while providing linguistic interpretations across speakers and spoofing conditions.
\end{abstract}

\begin{IEEEkeywords}
deepfake detection, speech processing, explainable artificial intelligence
\end{IEEEkeywords}


\section{Introduction}

As progress in the field of generative speech technology has pushed the deceptive capabilities of voice conversion and text-to-speech to greater levels, it is consequently ever more important to distinguish between bona fide and spoofed speech. Modern speech deepfake detection systems manage to achieve strong performance by leveraging self-supervised (SSL) speech representations extracted using large foundation models such as wav2vec 2.0~\cite{wav2vec}, HuBERT~\cite{hubert}, and WavLM~\cite{WavLM}. Despite these advances, the predictions of such systems remain largely difficult to understand with human-like reasoning~\cite{interpret} and they lack human-perceptible cues that would provide natural explanations for bona fide or spoof classifications. Even though a deepfake detector may successfully classify an utterance, it is often unclear which characteristics of the speech signal contributed to that prediction.

This lack of interpretability presents a challenge for the trustworthiness and transparency of speech deepfake detection systems. In applications where security and fairness are critical, understanding why a system reached a particular classification is crucial not only for faith in the decision itself, but also for future adjustment of the system. However, explanations based solely on low-level acoustic representations are often difficult to interpret, particularly for speech signals where the structure is more naturally described in linguistic terms, such as phonemes.

Phonemes provide a useful intermediate representation between raw acoustics and human interpretation of spoken language. In accordance with the International Phonetic Association and leading scholars, the definition of a phoneme is strictly limited to the assigned categorical representation of an uttered sound in a given language which does not incorporate its acoustic realization or varied pronunciation across dialects and spontaneous speech~\cite{phoneme}. Accordingly, we use 'phoneme' only when referring to the language-specific categorical label and 'phone' when referring to measurable acoustic properties or realizations.

Prior research has shown that spoofed speech exhibits deviations in a range of phonetic and prosodic characteristics, including segment durations~\cite{fujita}, non-speech regions~\cite{zhang}, and timing structure~\cite{tempdyn}. Nevertheless, it remains unclear to what extent a modern deepfake detector relies on these cues and whether particular phonemes (e.g.\ /o/, /s/, etc.) or phonetic categories (e.g.\ vowels, fricatives, etc.) contribute more strongly than others to model predictions. Furthermore, the contributions of such patterns across different spoofing attacks or different speakers could be better elucidated.

We introduce a phoneme-level framework for explainable speech deepfake detection alongside analysis of trends exposed by application of our framework to spoof/deepfake utterances in the ASVspoof~5 database~\cite{asvspoof5}. We combine a self-supervised front-end, a convolutional neural network (CNN) classifier, and a post-hoc explanatory module based on Gradient-weighted Class Activation Mapping (Grad-CAM)~\cite{Grad-CAM}. By aligning saliency maps with phoneme labels obtained through automatic speech recognition and forced alignment, model activations can be analyzed with respect to linguistically meaningful units. Through the aggregation of phoneme-level attributions across a large evaluation corpus, we investigate how detector behavior varies across spoofing attacks, speakers, and phonetic features and categories.

Experiments are conducted using the ASVspoof~5 corpus and a WavLM-based~\cite{WavLM} detector that achieves a competitive pooled equal error rate for the evaluation set~\cite{asvspoof5}. Statistical analysis reveals significant attack-dependent and speaker-dependent differences in phoneme-level activation patterns, with particularly strong effects observed for several vowels, fricatives, and silence or non-speech regions. These findings suggest that phoneme-level explanations can provide insight into the linguistic cues exploited by the analyzed deepfake detector.

The main contributions of this work are threefold:

\begin{itemize}
\item we introduce a phoneme-level explainability framework that links speech deepfake detector predictions to linguistically interpretable units;
\item we provide a large-scale statistical analysis of phoneme-level attributions across spoofing attacks and speakers;
\item we demonstrate that meaningful phonetic patterns can be extracted from a self-supervised speech deepfake detection system without sacrificing detection performance.
\end{itemize}

\section{Related Work}

We review related work in three areas relevant to the proposed framework: speech deepfake detection, explainability methods for speech processing, and phonetic analyses of deepfake speech. Together, these areas motivate the need for explanations that connect detector behavior to linguistically meaningful speech units.

\subsection{Speech Deepfake Detection}

The danger posed by neural speech synthesis and voice conversion systems has led to growing interest in and reliance on speech deepfake detection. Early approaches utilized hand-crafted acoustic features designed to capture artifacts introduced by synthesis and conversion algorithms~\cite{features, review}. More recently, deep learning systems based on CNNs and transformer architectures have become dominant~\cite{review}, particularly when paired with self-supervised speech representations learned from large quantities of unlabeled speech data~\cite{wangssl, hemlatassl, newssl, asvspoof5}.

Among these approaches, representations extracted from models such as wav2vec~2.0~\cite{wav2vec}, HuBERT~\cite{hubert}, and WavLM~\cite{WavLM} have demonstrated strong performance across a range of spoof/deepfake detection benchmarks~\cite{newssl, benchmarks}. These representations capture rich acoustic and linguistic information and often outperform systems trained using hand-crafted features~\cite{review, asvspoof5}. While detection accuracy has improved substantially, understanding what information these systems exploit remains an open challenge. In many cases, modern detectors achieve strong performance but provide little insight into which aspects of speech distinguish bona fide from spoofed utterances.

\subsection{Explainability in Speech Processing}

Explainable artificial intelligence aims to make model predictions more transparent by identifying characteristics or regions of an input that contribute to the output, i.e.\ a bona fide vs spoof prediction. In computer vision, saliency-based techniques~\cite{RISE} such as Grad-CAM~\cite{Grad-CAM} have become widely used for interpreting CNNs~\cite{XAIReview}. Similar approaches have been applied to text~\cite{LIME} and speech processing tasks including speech recognition~\cite{asr}, speaker verification~\cite{asv}, emotion recognition~\cite{emotion}, and deepfake detection~\cite{wanyingXAI}.

Despite their popularity, saliency methods present several challenges~\cite{XAIReview}. Highlighted regions indicate \textit{where} a model is sensitive but cannot directly explain \textit{what linguistic evidence} supports a model's prediction in meaningful terms that humans naturally use to interpret speech. Furthermore, Grad-CAM depends on intermediate convolutional representations. As such, the resulting explanations may reflect correlated features, such as segment durations or pitch contours, and can be sensitive to model architecture~\cite{adebayo2018sanity}, misleading in certain settings~\cite{attentionisnotexplanation}, and susceptible to adversarial manipulation~\cite{gradcamlies}. While existing work focuses on visualizing important time-frequency regions, relatively little effort has been devoted to translating these activations into linguistic units that can be readily understood by human analysts~\cite{manasi, liu24}.

\subsection{Phonetic Analysis of Deepfake Speech}

Generative speech technologies have long been known to exhibit phonetic and prosodic deviations relative to natural speech. Prior studies have reported differences in segment duration, coarticulation, speaking rate, pause structure, and articulatory consistency~\cite{SyntheticSpeech, tts87, forensic}. Although many of these artifacts have become less pronounced as synthesis quality has improved~\cite{evaltts}, they continue to provide valuable information for both forensic analysis and automatic detection systems~\cite{forensic}.

Several studies have investigated phoneme-level characteristics of spoofed speech, often focusing on acoustic measures such as phone duration or spectral properties~\cite{phonemeasv, SpeakerIndividuality, manasi, phonscore, phonemess}. However, relatively little work has examined whether modern deepfake detectors rely on these phonetic cues to help distinguish bona fide from spoofed speech. Consequently, there remains a gap between phonetic analyses of spoofed speech and explainability methods for deepfake detection. With the framework proposed in this work, we seek to bridge this gap by linking detector activations to phoneme-aligned segments and analyze them at scale.

\section{Phoneme-Level Explainability Framework}

Our framework comprises three components: an SSL-based front-end, a back-end deepfake detection classifier, and a phoneme-level explanatory module. Detection components produce scores for both bona fide and spoof classes, while the explanatory module aligns class-specific attribution signals with phonemes, enabling the analysis of detector behavior in linguistically interpretable terms. 

\subsection{Deepfake Detection System}

Input speech is first processed using an SSL-based front-end feature extractor, which produces frame-level representations encoding acoustic and linguistic information. These representations are then passed to a temporal CNN, which acts as the back-end classifier. The temporal frame-level outputs are aggregated using masked temporal average pooling, which averages the representations over speech frames while excluding padded frames introduced for batching. This produces a fixed-dimension, utterance-level representation. A final classification layer produces two logits corresponding to the bona fide and spoof classes.

While the detection architecture itself is not the primary contribution of this work, it provides a suitable deepfake detection system whose internal representations can be analyzed through the proposed explanatory framework. This allows us to investigate which characteristics of the speech signal contribute to the bona fide/spoof class scores and how these attribution patterns vary across speakers and spoofing attacks.




\subsection{Grad-CAM-Based Attribution}

To identify which components of an utterance provide support for each class hypothesis ('bona fide' or 'spoof'), Grad-CAM~\cite{Grad-CAM} is applied to the final convolutional layer of the classifier. Grad-CAM weights the activation maps of a convolutional layer by the gradients of a target output logit, thereby highlighting temporal regions of the input representation that are most associated with that target prediction. In this work, we compute separate Grad-CAM maps for the bona fide and spoof logits and refer to these class-specific attribution maps as \textit{bona fide-CAM} and \textit{spoof-CAM}, respectively.

The resulting attribution maps provide temporal evidence in support of each class hypothesis; both bona fide-CAM and spoof-CAM can exhibit non-zero attribution values for the same utterance, regardless of its ground-truth label. Rather than treating these maps as the final explanation, we use them as intermediate attribution signals that are subsequently mapped onto phoneme-labeled time spans.

\begin{figure}
\includegraphics[width=0.5\textwidth]{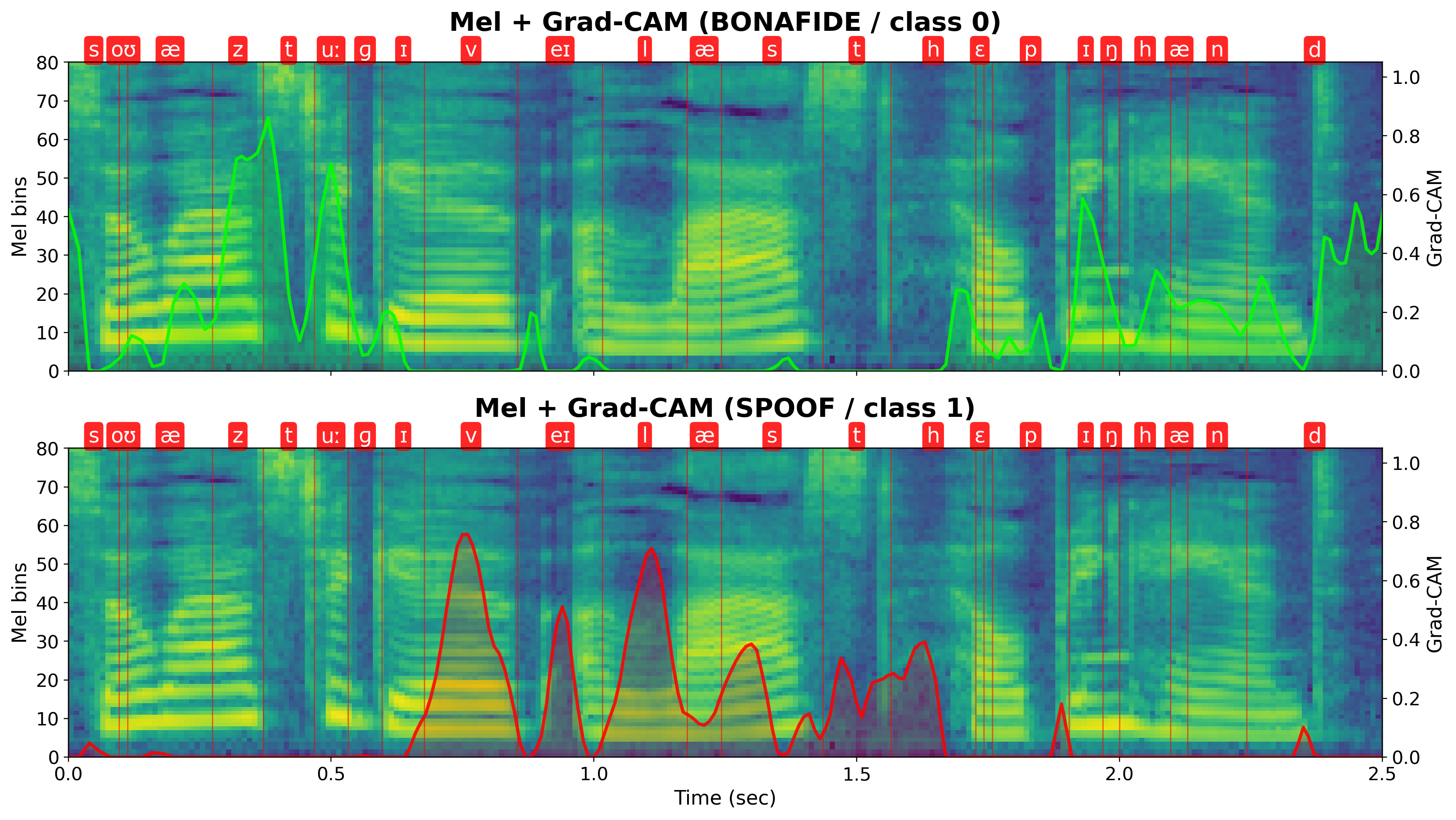}
\caption{Example of phoneme-aligned Grad-CAM attribution maps for the bona fide and spoof class hypotheses. 
The mel spectrogram is shown in the background, while the overlaid curves indicate the temporal Grad-CAM attribution scores for the bona fide logit (top) and spoof logit (bottom). 
Phoneme labels obtained from forced alignment are shown above each spectrogram. 
The two maps provide class-specific attribution signals and are therefore not mutually exclusive.
}
\label{fig:gradcam}
\end{figure}




\subsection{Phoneme Alignment and Attribution Aggregation}

To associate class-specific attribution scores with speech units, each utterance is first transcribed using Whisper~\cite{whisper} Turbo\footnote{\url{https://github.com/openai/whisper}}. The resulting transcript is then aligned to the acoustic signal using the Bournemouth Forced Aligner~\cite{bfa, CUPE}\footnote{\url{https://huggingface.co/Tabahi/CUPE-2i}} English model, which provides phoneme-level labels and corresponding phone boundaries.

Figure~\ref{fig:gradcam} shows an example of the output bona fide-CAM and spoof-CAM aligned with phonemes. The figure illustrates how the two class-specific attribution signals can highlight different temporal regions of the same utterance and how these regions can be associated with phoneme-labeled segments.

Given a phone or non-speech segment spanning frames $t_1$ through $t_2$, an importance score is computed by averaging the corresponding Grad-CAM attribution values within the segment. This produces one bona fide-CAM and one spoof-CAM importance score for each phone or non-speech segment in the utterance. By aggregating these scores across the corpus, we can examine how support for bona fide or spoof predictions varies across phoneme identities, phonetic categories, speakers, and spoofing attacks.




\subsection{Statistical Analysis}

Statistical analyses are performed on the aggregated phoneme-level attribution scores, which correspond to the average bona fide-CAM or spoof-CAM value within the aligned segment. We then analyze how the resulting score distributions vary across spoofing attacks, speakers, and phonetic categories.

For each phoneme, Kruskal-Wallis tests~\cite{kruskal} are used to assess whether the corresponding attribution scores differ significantly between spoofing attacks. This non-parametric test is used because the attribution scores are not assumed to follow a normal distribution. When a significant effect is observed, post-hoc pairwise comparisons are performed between attack pairs, with Benjamini-Hochberg correction applied to control for multiple comparisons~\cite{benjamini}.

In addition to statistical significance, we report epsilon-squared effect sizes, which range from 0 to 1 where $\varepsilon^2 \ge 0.01$ is small in magnitude, $\varepsilon^2 \ge 0.06$ is medium, and $\varepsilon^2 \ge 0.14$ is large~\cite{effectsize}. This is important because the large number of phoneme instances in the evaluation set can make even very small differences statistically significant. Effect sizes therefore help distinguish differences that are merely detectable due to sample size from differences that explain a meaningful proportion of the variation in attribution scores.

Additional analyses are performed at the phonetic-category level by grouping phonemes into vowels, fricatives, stops, nasals, affricates, approximants, and non-speech regions. For each utterance or attack condition, we identify which phonetic categories receive the largest attribution scores and use chi-square tests to examine whether the dominant categories are associated with the spoofing attack. This allows us to determine whether different attacks tend to concentrate class-specific support on different types of speech units.

\section{Experimental Setup}

Here, we describe the experimental protocol used to evaluate the proposed framework. We first outline our use of the ASVspoof~5 dataset and the partitions used for training and evaluation. We then detail our implementation, which is available as open-source for reproducibility under the same GPU environment\footnote{\url{https://github.com/eurecom-asp/PhonemeXAI}}, followed by the metrics used to assess detection performance and phoneme-level explainability.


\subsection{Dataset}

Experiments are conducted using the ASVspoof~5~\cite{asvspoof5} corpus, which contains bona fide speech and spoofed speech generated using a diverse collection of text-to-speech (TTS) and voice conversion (VC) systems. The dataset includes substantial variability in speaker characteristics, recording conditions, and spoofing attacks, making it suitable for both detection evaluation and interpretability analysis.


The official training partition is used for model optimization, while the ASVspoof~5 evaluation set is used for inference and statistical analysis. The training partition contains utterances collected from 400 speakers, spoofed utterances generated using 8 different attack algorithms, and a total of 18,797 bona fide and 163,560 spoofed utterances. The speaker-disjoint evaluation set contains data collected from 737 previously unseen speakers, spoofed utterances generated using 16 different attack algorithms, and a total of 680,774 utterances, of which 138,688 are bona fide and 542,086 are spoofed.

\subsection{Implementation Details}

In our experimental implementation, the SSL-based front-end is instantiated using WavLM~\cite{WavLM} Base+\footnote{\url{https://huggingface.co/microsoft/wavlm-base-plus}}, which is kept frozen during training. The temporal convolutional back-end consists of a single one-dimensional convolutional block with 512 channels, followed by masked temporal average pooling and a linear classification layer. The detector is trained using cross-entropy loss on the ASVspoof~5 training partition. Optimization is performed using the Adam optimizer~\cite{adam} for 200 epochs with a cosine annealing learning-rate schedule and warmup. The trained model is then applied to the ASVspoof~5 evaluation set, where both detection scores and intermediate activations are extracted for the subsequent explainability analysis.




\subsection{Evaluation Metrics}

Detection performance is measured using the Equal Error Rate (EER). Results are reported both as pooled EER and on a per-attack basis in order to characterize variability across spoofing conditions.

For explainability analysis, we evaluate the phoneme-level attribution scores by testing whether they vary systematically across spoofing attacks, speakers, and phonetic categories. Statistical significance tests are used to determine whether observed differences in attribution scores are unlikely to arise by chance, while effect-size measures are used to quantify the magnitude of these differences. This allows us to identify which phonemes and phonetic categories receive consistently different levels of class-specific support across attacks and speakers.

\section{Results and Analysis}

In this section, we report the detection performance of the proposed system and analyze class-specific phoneme-level attribution scores obtained from bona fide-CAM and spoof-CAM. We examine variation across spoofing attacks, temporal locations, non-speech regions, and speakers.

\subsection{Detection Performance}

Our detector achieves a pooled EER of 8.52\% for the ASVspoof~5 evaluation set. Rather than proposing a new detection architecture and comparing with other systems, this work instead focuses on explaining model predictions through our phoneme-level interpretability framework. While pooled performance is competitive~\cite{asvspoof5}, substantial variability is observed across spoofing attacks. Most attacks yield EERs below 8\%, with some EERs even below 4\%; however, attacks A24 (in-house ASR-based VC~\cite{asvspoof5}) and A28 (pre-trained YourTTS~\cite{yourtts}) remain considerably more challenging, with EERs exceeding 20\%. This variability suggests that \textbf{different generative speech systems introduce distinct artifacts} and motivates the attack specific phoneme-level analysis presented in the following sections.


\subsection{Attack-Dependent Phoneme Importance}

To investigate whether the detection of different spoofing attacks relies on distinct phonetic cues, we analyze frequency-normalized, phoneme-level spoof-CAM attribution scores across attacks. Frequency normalization was applied to remove the confounding effect of phoneme occurrence rates in the English language, which naturally follow the frequency distribution of function words in English regardless of whether the calculations come from spoken or written language~\cite{frequency}. Therefore, we ensure that attribution differences reflect the relative importance of a phoneme to the detector rather than its frequency in the corpus.

For each phoneme, we performed a Kruskal-Wallis~\cite{kruskal} test across attack conditions, followed by Benjamini-Hochberg~\cite{b-hoch} correction for multiple comparisons. Effect sizes were quantified using epsilon-squared ($\varepsilon^2$)~\cite{effectsize}, which estimates the proportion of attribution variance explained by attack identity.

Of the top 25 phonemes with the highest spoof-CAM scores after frequency normalization (including non-speech represented by the SIL token), all exhibit statistically significant differences across attacks ($p_{\mathrm{BH}} < 0.001$), indicating that the importance assigned to a phoneme by the spoof detector depends on the spoof generation method. However, effect sizes varied substantially between individual phonemes.

The strongest attack-dependent effects were observed for /o\textipa{U}/ ($\varepsilon^2=0.135$), /\textipa{E}/ ($\varepsilon^2=0.115$), /\textipa{AI}/ ($\varepsilon^2=0.113$), /s/ ($\varepsilon^2=0.101$), and /f/ ($\varepsilon^2=0.099$) with the top 10 phonemes ranked by attack-dependent variation shown in Table \ref{tab:phoneme_attack_effects}.  For /o\textipa{U}/ as much as 13.5\% of the variation in frequency-normalized importance is explained by the spoofing attack used to generate the speech. Rather than concentrating on a single universal artifact, \textbf{statistically significant attack-dependent variation is distributed across multiple phonemes and phonetic categories in the evaluated detector.}

Interestingly, the largest effects were concentrated among vowels such as /o\textipa{U}/, /\textipa{E}/, and /\textipa{AI}/, as well as fricatives, particularly /s/, /f/, and /h/.  Here the evidence suggests that attack-specific artifacts arise most strongly in acoustically complex speech regions, which are known to be challenging for TTS and VC systems~\cite{forensic}. 

Furthermore, analysis of bona fide-CAM activations revealed that many phonemes show substantial attack-dependent variation in their bona fide importance scores. In particular, /\textipa{2}/ ($\varepsilon^2=0.113$), /\textipa{E}/ ($\varepsilon^2=0.110$), /\textipa{\*r}/ ($\varepsilon^2=0.108$), and /o\textipa{U}/ ($\varepsilon^2=0.107$) emerged as especially attack-dependent; more results are shown in Table \ref{tab:phoneme_attack_effects}. Considering the prominence of vowels here, the detector is likely sensitive to speaker-dependent articulatory variability arising more strongly in vowels and lightly constricted sounds like approximants when attempting to identify natural speech. Moreover, the large attack effects indicate that different spoofing systems distort these phonetic cues in distinct ways, resulting in attack-specific attribution patterns.

\begin{table}[t]
\centering
\caption{Phonemes exhibiting the strongest attack-dependent variation in frequency-normalized Grad-CAM attribution scores for both spoof and bona fide predictions, ranked by Kruskal--Wallis effect size ($\varepsilon^2$).}
\label{tab:phoneme_attack_effects}
\begin{tabular}{lc|lc}
\hline
\multicolumn{2}{c|}{Spoof-CAM} &
\multicolumn{2}{c}{Bona fide-CAM} \\
\hline
Phoneme & $\varepsilon^2$ &
Phoneme & $\varepsilon^2$ \\
\hline
/o\textipa{U}/      & 0.135 & /\textipa{2}/       & 0.113 \\
/\textipa{E}/       & 0.115 & /\textipa{E}/       & 0.110 \\
/\textipa{AI}/      & 0.113 & /\textipa{\*r}/      & 0.108 \\
/s/                 & 0.101 & /o\textipa{U}/      & 0.107 \\
/f/                 & 0.098 & /i\textipa{:}/      & 0.102 \\
/z/                 & 0.092 & /\textipa{I}/       & 0.099 \\
/e\textipa{I}/      & 0.091 & 
/h/                 & 0.097 \\
/i\textipa{:}/      & 0.088 & 
/v/                 & 0.097 \\
/h/                 & 0.086 & /e\textipa{I}/      & 0.096 \\
/w/                 & 0.072 & /\textipa{AI}/      & 0.091 \\
\hline
\end{tabular}
\end{table}

\subsection{Temporal Localization}


Beyond identifying which phonemes contribute to detector predictions, we also investigated the exact location within a phone and within the entire utterance where the highest Grad-CAM activations occur for both spoof and bona fide predictions. For each activation peak, we computed the temporal distance to the nearest phone boundary designated by the Bournemouth Forced Aligner and to the first and last 20\% of the utterance duration, and we find significant differences between human speech and the various spoofing attacks only for phone boundary distance.  

\begin{figure}[t]
\includegraphics[width=0.5\textwidth]{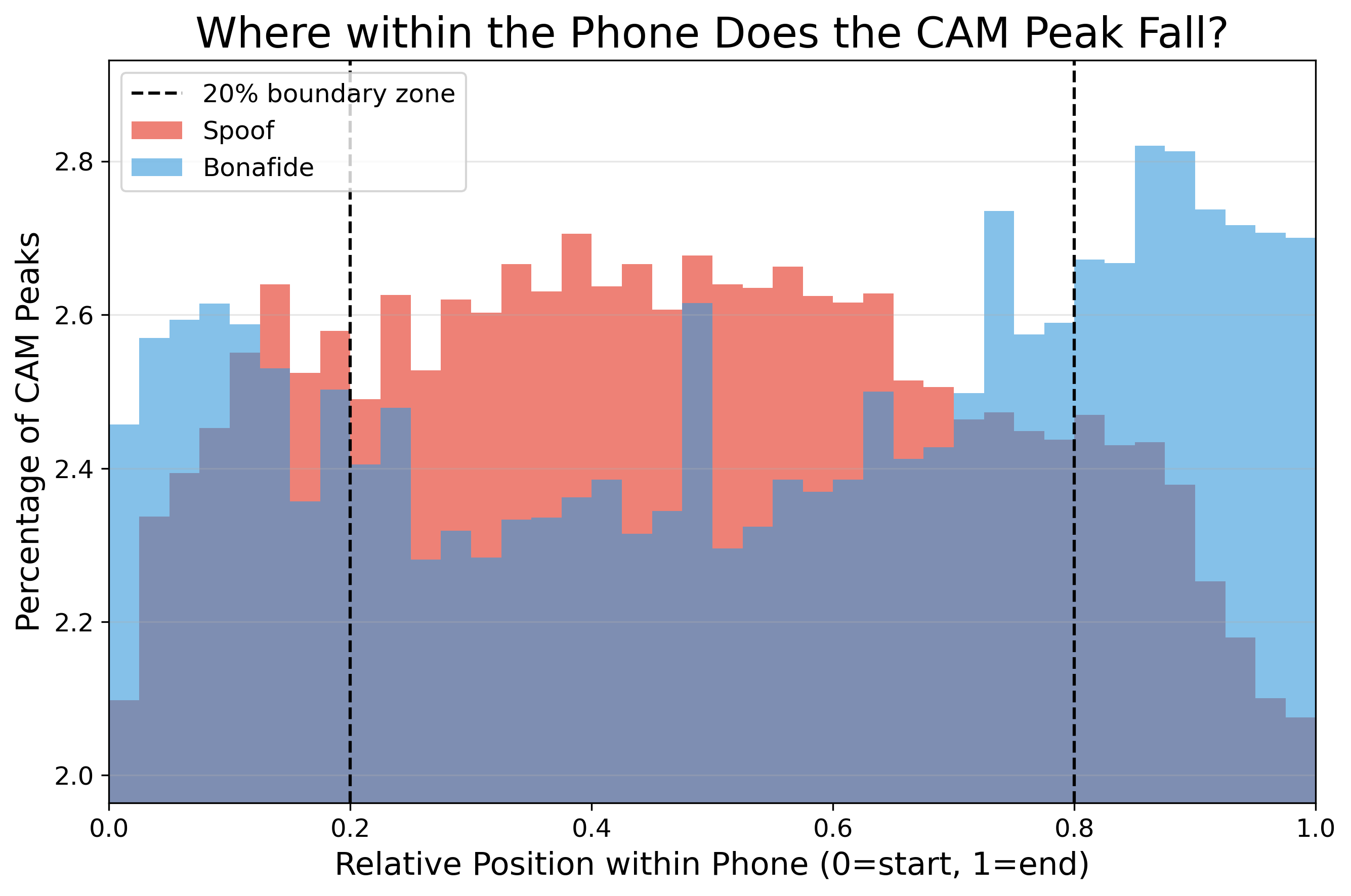}
\caption{Grad-CAM activation peaks differ in their relative location within the phone depending on the model's prediction as spoof or bona fide, representing over 2\% of the CAM peaks.}
\label{fig:boundary}
\end{figure}

Across all attacks, bona fide activation peaks were consistently concentrated near phone boundaries, representing the first or last 20\% of the duration, while the spoof activations concentrated within the phones, as seen in Figure~\ref{fig:boundary}. Median boundary distances ranged from approximately 10-15 ms, whereas median distances within phone interiors ranged from approximately 37-56 ms. This finding indicates that the evaluated detector relies on acoustic transitions between phonetic units for determining natural human speech and on steady-state phonetic realizations for detecting spoofed speech.

Attack-dependent differences were also substantial. Pairwise comparisons revealed significant differences for 124 of 136 attack pairs (16 spoofing attacks plus bona fide) after Benjamini--Hochberg correction ($p < 0.05$), indicating that \textbf{spoofing attacks differ not only in the phonemes that attract model attention but also in the temporal location of the salient evidence.}

\subsection{Non-Speech and Pause Regions}

Non-speech and pause regions are consistently found to be informative for spoof predictions, and even more so for bona fide.  On average, between 7\% and 25\% of total Grad-CAM spoof and bona fide attribution mass was assigned to non-speech segments with significant differences across ASVspoof~5 attacks.  Attack identity had a particularly strong effect on bona fide-CAM non-speech attribution ($\varepsilon^2=0.102$), substantially exceeding the corresponding spoof-CAM effect ($\varepsilon^2=0.043$).  This is in line with other findings~\cite{silencegolden, silence2, chettri} that have focused specifically on non-speech regions to detect natural human speech phonation and pause patterns which generative speech technologies struggle to accurately replicate.  

The most extreme differences can be seen between attacks A20 (in-house unit-select TTS~\cite{asvspoof5} and Malafide adversarial attack~\cite{malafide}) and A21 (ToucanTTS~\cite{toucan} and BigVGAN~\cite{bigvgan}).  Despite both systems using TTS with similar detection performance (5.5\% EER vs 6.5\%), the former system has approximately 25\% of the bona fide attribution mass and 21\% of the spoof-CAM concentrated in non-speech regions, whereas, for the latter, non-speech content represents only 7\% of bona fide attributions and 8\% of spoof.  This contrast suggests that \textbf{our deepfake detector unevenly exploits attack-specific characteristics that may vary widely} such as pause duration, pause placement, turn-taking timing, voicing onset/offset behavior, and transitions between speech and non-speech segments.   

\subsection{Speaker-Dependent Effects}

\begin{figure}[t]
\includegraphics[width=0.5\textwidth]{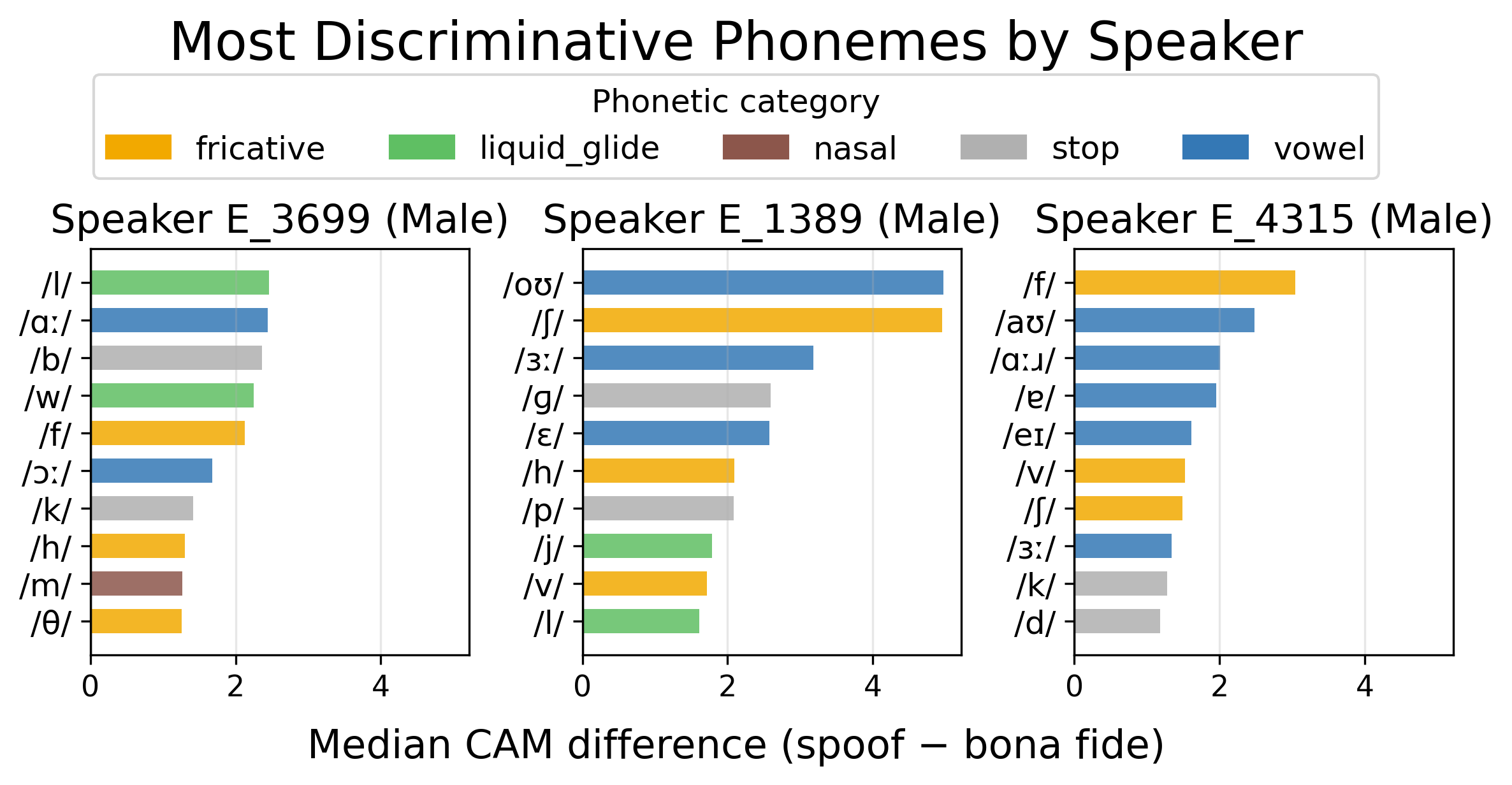}
\caption{Example of the most discriminative phonemes for three representative speakers. Bars show the median difference between spoof-CAM and bona fide-CAM attribution scores for each phoneme, grouped by phonetic category. The relative importance of phonemes differs substantially across speakers, indicating that no single phoneme consistently dominates across all speakers.}
\label{fig:speakers}
\end{figure}

In addition to attack-dependent variation, substantial speaker-dependent effects are also observed. Detector behavior varies considerably at the speaker level, both in terms of decision confidence and the phonetic evidence used to support spoof predictions. Figure \ref{fig:speakers} illustrates the highest-ranking discriminative phonemes for a subset of three speakers, revealing that the phonemes receiving the greatest Grad-CAM attribution differ markedly across individuals. For example, /l/, /\textipa{A:}/, and /b/ are among the most discriminative phonemes for speaker E\_3699, whereas /o\textipa{U}/, /\textipa{S}/, and /\textipa{3:}/ dominate for speaker E\_1389, and /f/, /a\textipa{U}/, and /\textipa{A:\*r}/ for speaker E\_4315. These differences indicate that \textbf{the analyzed detector does not rely on a fixed set of universally informative phonemes but instead adapts to speaker-specific acoustic characteristics that reveal substantial variation in per-speaker vulnerability.}

Speaker characteristics also influence broader phonetic attribution patterns. Significant gender-dependent differences in spoof probability were observed for 15 of the 16 spoofing attacks after Benjamini-Hochberg correction, indicating that the detector's confidence depends on speaker gender for nearly all synthesis conditions except MaryTTS~\cite{marytts}. Likewise, the proportion of spoof-CAM assigned to voiced phones differed significantly between genders for 12 attacks, while non-speech and pause regions exhibited significant gender effects for 13 attacks. Bona fide speech also showed significant gender effects for both voiced and non-speech attribution distributions. Together, these findings suggest that detector predictions are affected not only by synthesis artifacts themselves but also by how those artifacts are shaped by speaker identity. Consequently,  \textbf{attribution patterns vary jointly by spoofing attack and speaker characteristics, with the phonetic evidence for a given attack varying systematically across speakers,} although our speaker analysis reflects the evaluation corpus rather than a balanced speaker population.

\section{Discussion}

Our analysis reveals significant variation in class-specific attribution patterns across spoofing attacks, speakers, and phonetic information. Vowels, fricatives, approximants, non-speech regions, voicing, and speaker gender emerge as particularly informative, while the relative importance of individual phonemes and the temporal location within phones varies substantially across attacks and speakers.

\subsection{Limitations and Future Work}

The proposed framework extends conventional saliency analysis by transforming Grad-CAM activations into statistically-validated, phoneme-level interpretations, providing meaningful insight into the factors influencing the analyzed detector's predictions.  Notably, Grad-CAM itself does not establish direct causal attribution, and it is possible that highlighted phonemes may reflect correlated acoustic properties. In addition, explanations are conditioned on the underlying network architecture and may differ across alternative detector designs.

The framework also depends on automatic transcription and forced alignment. Errors introduced during either stage may affect phone boundaries and attribution estimates, particularly for highly degraded or strongly synthetic speech.

Future work will investigate alternative attribution techniques, incorporate prosodic measures, and extend the analysis to multilingual settings, cross-architecture validation, human evaluation, and phoneme-ablation tests. An important direction will be determining whether the phonemes identified as salient correspond to measurable acoustic deviations in deepfake speech, thereby strengthening the connection between model explanations and phonetic phenomena.


\section{Conclusions}


After applying our framework to ASVspoof~5 data, the evidence shows significant speaker-dependent effects as well as attack-dependent patterns across both phonetic features (voicing, pauses, etc.) and phoneme classes (vowels, fricatives, etc.). These findings reinforce the claim that the evaluated detector can be systematically analyzed in linguistically interpretable terms without compromising performance. Overall, this work demonstrates that a phoneme-level explainability framework can transform low-level saliency into interpretable insights for systematic analysis of how a deepfake detector behaves across speakers and spoofing conditions.

\section{AI Use Disclosure}

During this research, Generative AI tools (OpenAI ChatGPT and Anthropic Claude) were used to assist with manuscript formatting and editing as well as development of analysis code. The authors reviewed, revised, and verified all generated content and accept full responsibility for the accuracy and originality of the work.

\bibliographystyle{IEEEtran}
\bibliography{slt}

\end{document}